\documentclass{article}
\usepackage{fullpage,graphicx,natbib,setspace,subcaption,authblk,hyperref,doi}

\hypersetup{
  colorlinks=true,
  linkcolor=blue,          % color of internal links (change box color with linkbordercolor)
  citecolor=blue,        % color of links to bibliography
  filecolor=blue,      % color of file links
  urlcolor=blue           % color of external links
}
\title{A C++ template library for efficient forward-time population genetic simulation of large populations}
\author{Kevin R. Thornton}
\affil{Department of Ecology and Evolutionary Biology\\
University of California, Irvine\\
\href{mailto:krthornt@uci.edu}{krthornt@uci.edu}}

\begin{document}
\maketitle

\section*{Abstract}
\texttt{fwdpp} is a C++ library of routines intended to facilitate the development of forward-time simulations under arbitrary mutation and fitness models.  The library design provides a combination of speed, low memory overhead, and modeling flexibility not currently available from other forward simulation tools. The library is particularly useful when the simulation of large populations is required, as programs implemented using the library are much more efficient than other available forward simulation programs.

\section*{Introduction}
The last several years have seen an increased interest in simulating populations forwards in time \citep{Messer:2013ct,Hernandez:2008kn,Peng:2008em,Peng:2007fx,Pinelli:2012fc,Padhukasahasram:2008ee,ChadeauHyam:2008gj,CarvajalRodriguez:2008gz,Peng:2010bi,Kessner:ip,Neuenschwander:2008bf} in order to understand models with natural selection at multiple linked sites which cannot be easily treated using coalescent approaches.  Compared to coalescent simulations, forward-time simulations are extremely computationally intensive, and several early efforts may not be efficient enough for in-depth simulation studies (reviewed in \cite{Messer:2013ct}).  More recently, two programs, \texttt{sfs\_code} \citep{Hernandez:2008kn} and \texttt{SLiM} \citep{Messer:2013ct} have been introduced and demonstrated to be efficient enough (both in run-time and memory requirements) to obtain large numbers of replicates, at least for the case of simulating relatively small populations.  Both of these programs are similar in spirit to the widely-used coalescent simulation program \texttt{ms} \citep{Hudson:2002vy} in that they attempt to provide a single interface to simulating a vast number of possible demographic scenarios while also allowing for multiple selected mutations, which is not possible on a coalescent framework.  The intent of both programs is to allow efficient forward simulation of regions with large scaled mutation and recombination rates ($\theta=4N\mu$ and $\rho=4Nr$, respectively, where $N$ is the number of diploids, $\mu$ is the mutation rate per gamete per generation, and $r$ is the recombination rate per diploid per generation) by simulating a relatively small $N$ and relatively large $\mu$ and $r$ \cite[also see][for another example of a similar strategy]{Hoggart:2007kp,ChadeauHyam:2008gj}.  This ``small $N$'' strategy allows a sample of size $n \ll N$ to be taken from the population in order to study the effects of complex models of natural selection and demography on patterns of variation in large chromosomal regions.  \cite{Messer:2013ct} has recently shown that his program \texttt{SLiM} is faster than \texttt{sfs\_code} for such applications and requires less memory.  However, both programs are efficient enough such that either could be used for the purpose of investigating the properties of relatively small samples.

The modern era of population genomics involving large samples \citep{GenomesProjectConsortium:2010gj,McVean:2012co,Pool:2012cx,Cao:2011cfa,Mackay:2012fda} and very large association studies in human genetics \citep{Burton:2007ht} demonstrate a need for efficient simulation methods for relatively large population sizes.  For example, simulating current human GWAS with thousands of individuals would require simulating a population much larger than the number of cases plus controls. Further, the simulation of complex genotype-to-phenotype relationships will require parameters such as random effects on phenotype and fitness (not currently implemented in \texttt{SLiM} nor in \texttt{sfs\_code}) such that heritability is less than one \citep[see][for existing examples of such simulations]{Kessner:ip,Neuenschwander:2008bf,Peng:2008em,Pinelli:2012fc,thornton:2013dl}. 

In this article I present \texttt{fwdpp}, which is a C++ library for facilitating the implementation of forward-time population genetic simulations.  Rather than attempt to provide a general program capable of simulating a wide array of models under standard modeling assumptions akin to \texttt{ms}, \texttt{SLiM}, or \texttt{sfs\_code}, \texttt{fwdpp} instead abstracts the fundamental operations required for implementing a forward simulation under custom models.  An early version of the code base behind \texttt{fwdpp} has already been used successfully to simulate a novel disease model in large population that would not be possible with existing forward simulations \citep{thornton:2013dl} and to simulate ``evolve and resequence'' experiments such as \citep{Burke:2010eq,BaldwinBrown:2014eaa}.  Since the publication of those papers, the library code has been improved in many ways, reducing run times by more than a factor of two.  \texttt{fwdpp} provides a generic interface to procedures such as sampling gametes proportional to their marginal fitnesses, mutation, recombination, and migration.  The use of advanced C++ techniques involving code templates allows a library user to rapidly develop novel forward simulations under any mutation model or fitness model (including disease models as discussed above).  The library is compatible with another widely-used C++ library for population genetic analysis \citep[\texttt{libsequence,}][]{Thornton:2003vz} and contains functions for generating output compatible with existing programs based on \texttt{libsequence} for calculating summary statistics.  Further, the run-time performance of programs implemented using \texttt{fwdpp} compare quite favorably to \texttt{SLiM} for the ``small $N$'' case described above.  However, for the case of large $N$, \texttt{fwdpp} results in programs with significantly smaller run times and memory requirements then either \texttt{SLiM} or \texttt{sfs\_code}, allowing for very efficient simulation of samples taken from large populations for the purposes of modeling population genomic data sets or large case/control studies.

\section*{Sampling algorithm}
The library supports two sampling algorithms for forward simulation.  The first of these is an individual-based method, where $N$ diploids are represented.  Descendants are generated by sampling parents proportionally to their fitnesses, followed by mutating and recombining the parental gametes.  Below, I show that the individual-based method results in the fastest run time for models involving natural selection.  Therefore, for most applications, the individual-based sampling functions should be considered the default choice for developing custom simulations.

The second algorithm is gamete-based.  In this algorithm, no diploids are represented.  Rather, in any generation $t$, there are $g_t$ gametes, each with $0 < x < 2N$ copies present in the population.  In order to generate the next generation, the expected frequency of each gamete in the next generation is obtained using the formula
\[
p'_i = \frac{p_iw_i}{\bar{w}},
\]
where $p'_i$ is the expected frequency of the $i^{th}$ gamete in the next generation, $p_i$ is its current frequency ($\frac{x}{2N}$), and $w_i = \frac{\sum_{j=1}^{j=g_t}P_{ij}w_{ij}}{p_i}$ is the marginal fitness of the gamete over all possible diploid genotypes ($P_{ij}$) containing the $i^{th}$ gamete \citep[p.  179]{Crow:1971wa}.  The expected frequencies of each gametes are used in one round of multinomial sampling to obtain the number of copies of each gamete in the next generation.  Although slower than the individual-based sampler for models with selected mutations, the gamete-based sampler reflects the original code base of \texttt{fwdpp}, previously used in \citep{thornton:2013dl,BaldwinBrown:2014eaa}.  This code provides only one additional function to the library user and requires fewer data structures (as no container of diploids is needed).  It is therefore kept in the library both for backwards compatibility with previous projects and for the possibility of future performance improvements. 

\section*{Library design}
\begin{figure}[!h]
  \centering
  \includegraphics[scale=0.5]{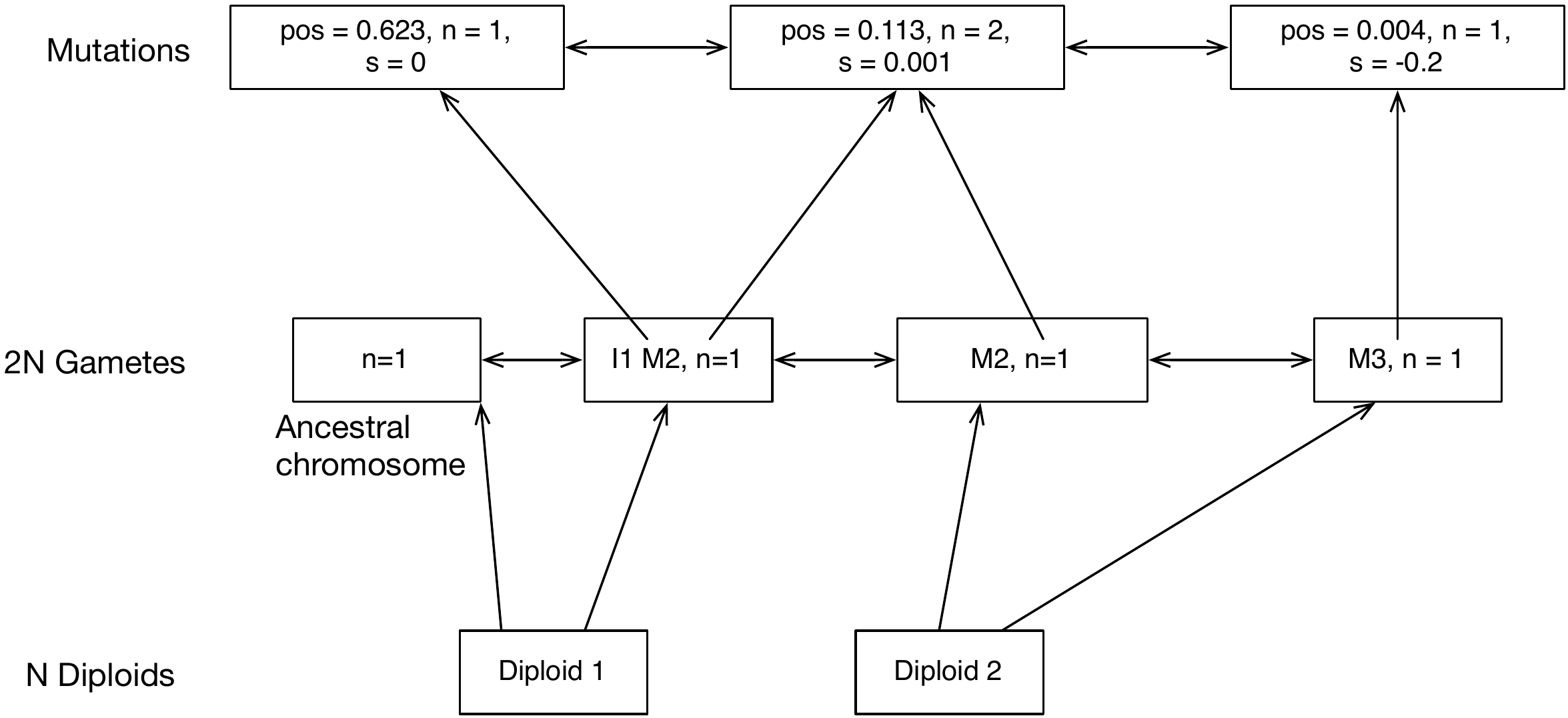}
  \caption{\label{fig:schemaind} Major data structures used by the simulation library for individual-based simulation.  Mutations are stored in a doubly-linked list.  Within the list, each mutation occupies a unique place in memory accessible via a C++ pointer. The pointers to the three mutations are labelled M1, M2, and M3.  Gametes are containers of pointers, meaning that the data for any specific mutation is stored only once and may be accessed via the pointers contained by gametes bearing that mutation.  The ``gamete pool'' of a population is also stored in a doubly-linked list. The entire population is thus represented by three data structures: a list of mutations, a list of gametes containing pointers into the mutation list, and a vector of diploids.}
\end{figure}

The intent of the library is to provide generic routines for mutation, recombination, migration and sampling gametes proportionally to their fitnesses in a finite population of $N$ diploids.  The library does this in a memory-efficient manner by defining a small number of simple data types.  First, there are mutations. The simplest mutation type is represented by a position and an integer representing its count in the population ($0 \leq n \leq 2N$).  Second, there are gametes, which are containers of pointers to mutations.  Finally, in individual-based simulations, there are diploids, which are pairs of pointers to gametes.  The schema relating these data structures is shown in Figure \ref{fig:schemaind}.  The details of the relations between data types in individual-based simulation are shown in Figure \ref{fig:uml}.  This pointer-based structure is perhaps obvious, but it has several advantages.  First, it replaces copying of data with copying of pointers, which is both faster and much more memory efficient.  Second, because each pointer is unique, we can ask if two gametes carry the same mutation by asking if they contain the same pointers, with no need to query the actual position, etc., of the mutation object pointed to. Finally, storing pointers to neutral and non-neutral mutations in separate containers typically speeds up the calculation of fitness because most models of interest will involve a relatively small proportion of selected mutations compared to the total amount of variation in the population. 

Library users create their own custom data types primarily by extending \texttt{fwdpp}'s built-in mutation type by creating a new mutation type that inherits from the built in type (described above), and adding the new required data. For example, selection coefficients, origination and fixation times, etc., may be tracked by a custom mutation type (Figure \ref{fig:uml}).  The gamete type is then a simple function of the custom mutation type and the container in which these mutations are stored (Figure \ref{fig:uml}).  

These user-defined data types are passed to functions implementing the various sampling algorithms required for the simulation.  Because the library cannot know ahead of time what the ``rules'' of the simulation are, library algorithms are implemented in terms of templates, which may be thought of as skeleton code for a particular algorithm. In other words, a template function could be implemented in terms of type ``T'', which could be an integer, floating-point number, or a custom data type as decided by the programmer using the function.   The substitution of specific types for the place holders (and related error-checking) is performed by the compiler. In standard C++, templates are used to implement algorithms on data stored in containers (such as sorting, \cite[pp. 94-101]{Josuttis:1999tx}).  The behavior of these algorithms may be modified by custom \textit{policies} \citep[pp. 119-134]{Josuttis:1999tx}.  For example, a sorting order may be affected by a policy.  Similarly, users of \texttt{fwdpp} provide policies specifying the biology of the population at each stage of the life cycle. An example of a policy function would be the mutation model.  A mutation model policy must specify the position and initial frequency of a new mutation along with any other data such as selection coefficients, dominance, etc.  Many of the most commonly-used policies for standard population genetic models (multiplicative fitness, how mutation containers are updated after sampling, etc.) are provided by the library.  A typical custom policy typically involves little new code, and the example programs distributed with the library demonstrate this point.  The library also comes with additional documentation detailing the concept of policies in standard C++ and how that concept is applied in \texttt{fwdpp} and what the minimal requirements are for each type of policy (mutation, migration, and fitness being the three most important).  The ability to extend the built-in mutation and gamete types and combine them with custom policies facilitates the implementation of algorithms for simulation under arbitrary models.  As the library has developed, I have found that it has evolved to a point where the balance between inheritance (the ability to build custom types from existing types, such as mutations) and template-based data types and functions is such that new models may be implemented with relatively little new code being written.

\section*{Library features}

The library contains several features to facilitate writing efficient simulations.  As of library version 0.2.0, these features are supported for both the gamete- and individual- based portions of \texttt{fwdpp} and include:
\begin{enumerate}
\item The ability to initialize a population from the output of a coalescent simulation stored in the format of the program \texttt{ms} \citep{Hudson:2002vy}.   This input  may come either from an external file or the coalescent simulation could be run internally to the program, for example using the routines in \texttt{libsequence} \citep{Thornton:2003vz}.  The routines are compatible with coalescent simulation output stored in binary format files using routines in \texttt{libsequence} version $\geq 1.7.8$. 
\item Samples from the population may be obtained in \texttt{ms} format.
\item The ability to copy the containers of mutations and gametes into new containers.  The result of the copy operation is an exact copy of the population which can be evolved independently.  Applications include simulating replicated experimental evolution \citep{BaldwinBrown:2014eaa} or conditioning simulation results on a desired event, such as the fate of a particular mutation, and the population state is repeatedly restored and evolved until the desired outcome is reached via naive rejection sampling.
\item The population may be written to a file in a compact binary format.  This binary output may then be used as input for later simulation.  Applications of this feature include storing populations simulated to equilibrium for later evolution under more complex models and/or storing the state of the population during the course of a long-running simulation such that it may be restarted from that point in case of unexpected interruptions. 

\end{enumerate}

\section*{Library dependencies}
The code in \texttt{fwdpp} uses the C-language GNU Scientific Library (``GSL'', \url{http://www.gnu.org/software/gsl/}) for random number generation.  The \texttt{boost} libraries (\url{http://www.boost.org}) are used extensively throughout the code.  Finally, \texttt{libsequence} \citep{Thornton:2003vz} was used to implement the input and output in \texttt{ms} format described in the previous section.  All three of these libraries must be installed on a user's system and be accessible to the system's C++ compiler.

\section*{Documentation and example programs}
The library functions are documented using the \texttt{doxygen} system (\url{http://www.doxygen.org}).  The documentation includes a tutorial on writing custom mutation and fitness functions.  The library also contains several example programs whose complete source codes are available in the documentation.  The simplest of these programs are \texttt{diploid} and \texttt{diploid\_ind},  which use the gamete- and individual-based methods, respectively, to simulate a population of $N$ diploids with mutation, recombination and drift and output a sample of size $0 < n \ll 2N$ in the same format as \texttt{ms} \citep{Hudson:2002vy}. The remaining example programs add complexity to the simulations and document the differences with respect to these programs.  All of the example programs model mutations according to the infinitely-many sites model \citep{Kimura:1969tn} with both the mutation and recombination rates being uniform along the sequence.  (Non-uniform recombination rates are trivial to implement via custom policies returning positions along the desired genetic map of the simulated region.)  In practice, I expect that future programs developed using \texttt{fwdpp} will use the individual-based sampler due to its speed in models with selection (see below).  Many of the examples are implemented using both the gamete- and individual- based sampling methods.  The names of source code files and binaries for the latter have the suffix ``\_ind'' added to them to highlight the difference.  

The complete library documentation and example code are distributed with the source code (see {\sc Availability} below).  All of the performance results described below are based on the example programs.

\section*{Availability}

\texttt{fwdpp} is release under the GNU General Public License (GPL, \url{http://www.gnu.org/licenses/gpl.html}).  The primary web page for all software from the author is \url{http://www.molpopgen.org/software/}, where links to the main \texttt{fwdpp} page may be found.  The source code is currently distributed from \url{https://github.com/molpopgen/fwdpp}.

\section*{Performance}
Performance under the constant-sized Wright-Fisher model without selection was evaluated using the UCI High Performance Computing cluster, which consists of dozens of 64-core nodes, mainly with AMD Opteron 6274 processors.  An entire queue of three such nodes was reserved for performance testing, ensuring that no disk-intensive processes were running alongside the simulations and degrading their performance.  All code was compiled using the GNU Compiler Collection (GCC) suite (\url{http://gcc.gnu.org}) version 4.7.2.  Programs based on \texttt{fwdpp} depended on \texttt{boost} version 0.5.3 (\url{http://www.boost.org}), \texttt{libsequence} version 1.7.8 (\url{http://www.molpopgen.org}) and the GNU Scientific Library (GSL, \url{http://gnu.org/software/gsl}) version 1.16.  The GSL v1.16 was also used to compile \texttt{SLiM} \citep{Messer:2013ct}.  The software versions used for all results were \texttt{fwdpp} version 0.2.0, \texttt{SLiM} version 1.7, and \texttt{sfs\_code} version 2013-07-25.   For all simulations, \texttt{sfs\_code} was run with the infinitely-many sites mutation option.

\begin{figure}[!h]
  \centering
  \includegraphics[scale=0.4]{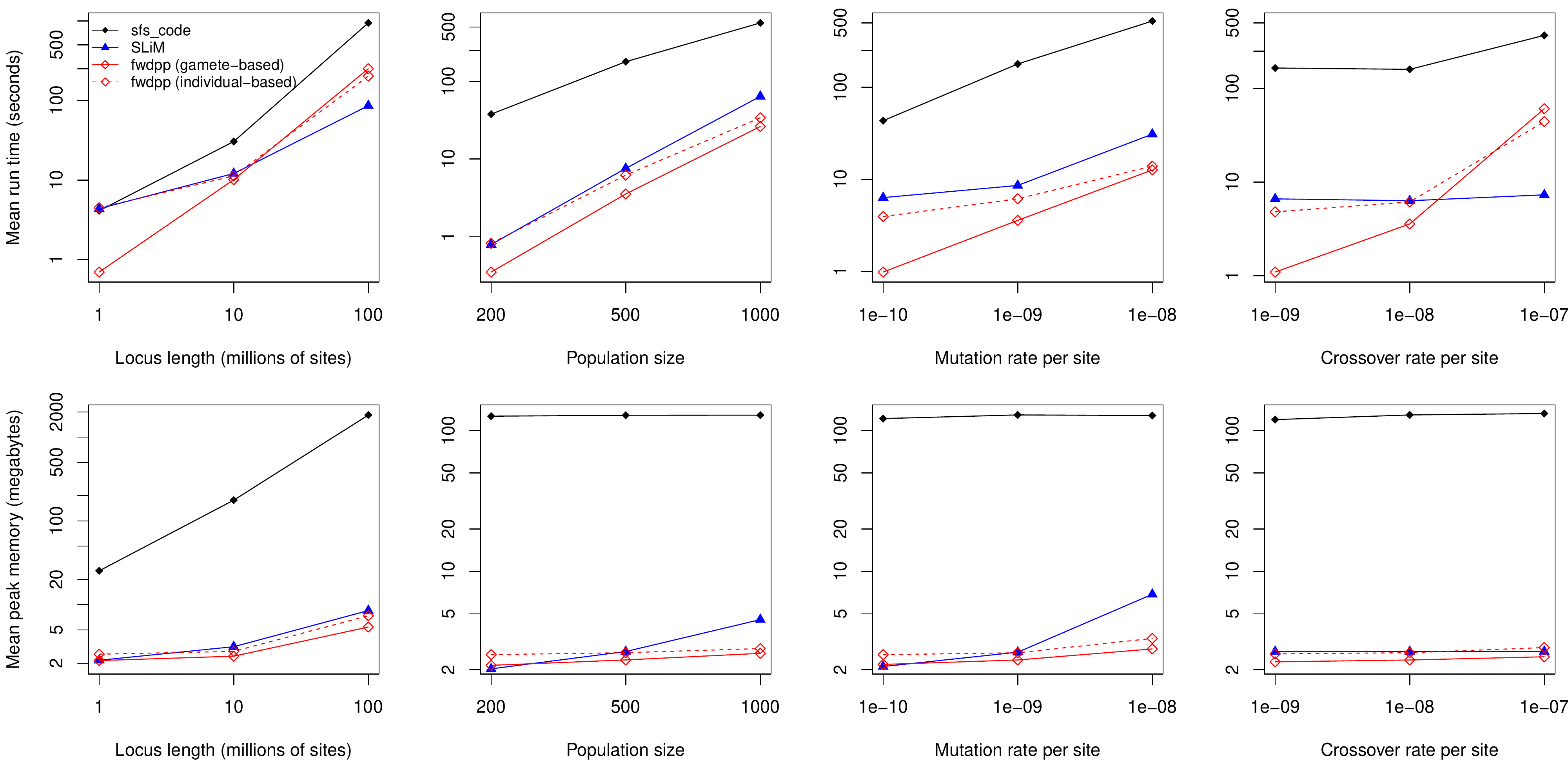}
  \caption{\label{fig:performance1} Performance comparison for the case of small population size ($N$).  Shown are the means of run time and of peak memory use for \texttt{sfs\_code}, \texttt{SLiM}, and a program written using \texttt{fwdpp}.  Note that the y-axis is on a log scale. The results are based on 100 simulations with the following base parameter values: diploid population size $N = 500$, locus length $L = 5 \times 10^6$ base pairs, mutation rate per site $\mu_{bp} = 1 \times 10^{-9}$, and recombination rate per diploid per site $r_{bp} = 1\times 10^{-8}$. (Both \texttt{SLiM} and \texttt{sfs\_code} parameterize per-generation rates as per-base pair.) All simulations were run for $10N$ generations.  For each column of figures, one of the four parameters was varied while the remainder were kept at their base values.  For the leftmost column, \texttt{sfs\_code} was run with 100 loci of length $L/100$ for all $L > 10^6$. Simulations implemented using \texttt{fwdpp} do not explicitly model sites and instead are implemented in terms of the usual scaled mutation and recombination rates $\theta=4NL\mu_{bp}$ and $\rho=4NLr_{bp}$, respectively.}
\end{figure}

Figure \ref{fig:performance1} shows the average run times and memory requirements of \texttt{sfs\_code} \citep{Hernandez:2008kn}, \texttt{SLiM} \citep{Messer:2013ct}, and \texttt{fwdpp} over a variety of parameter values where the population size, $N$, is small ($\le 1,000$).  For nearly all parameter combinations, \texttt{SLiM} and \texttt{fwdpp} are much faster than \texttt{sfs\_code} and require less memory.  When the total amount of recombination gets very large (either the locus length gets very long and/or the recombination rate gets large), \texttt{fwdpp} was slower than \texttt{SLiM} but still several times faster than \texttt{sfs\_code}.  Holding the population size and recombination rate constant, \texttt{fwdpp} was faster than \texttt{SLiM} as either the population size increases or the mutation rate increases (middle two columns of Figure \ref{fig:performance1}).  Although Figure \ref{fig:performance1} suggests very large relative differences in performance, it is important to note that the absolute run times are still rather short for all three programs. 

\begin{figure}[!h]
  \centering
  \includegraphics[scale=0.25]{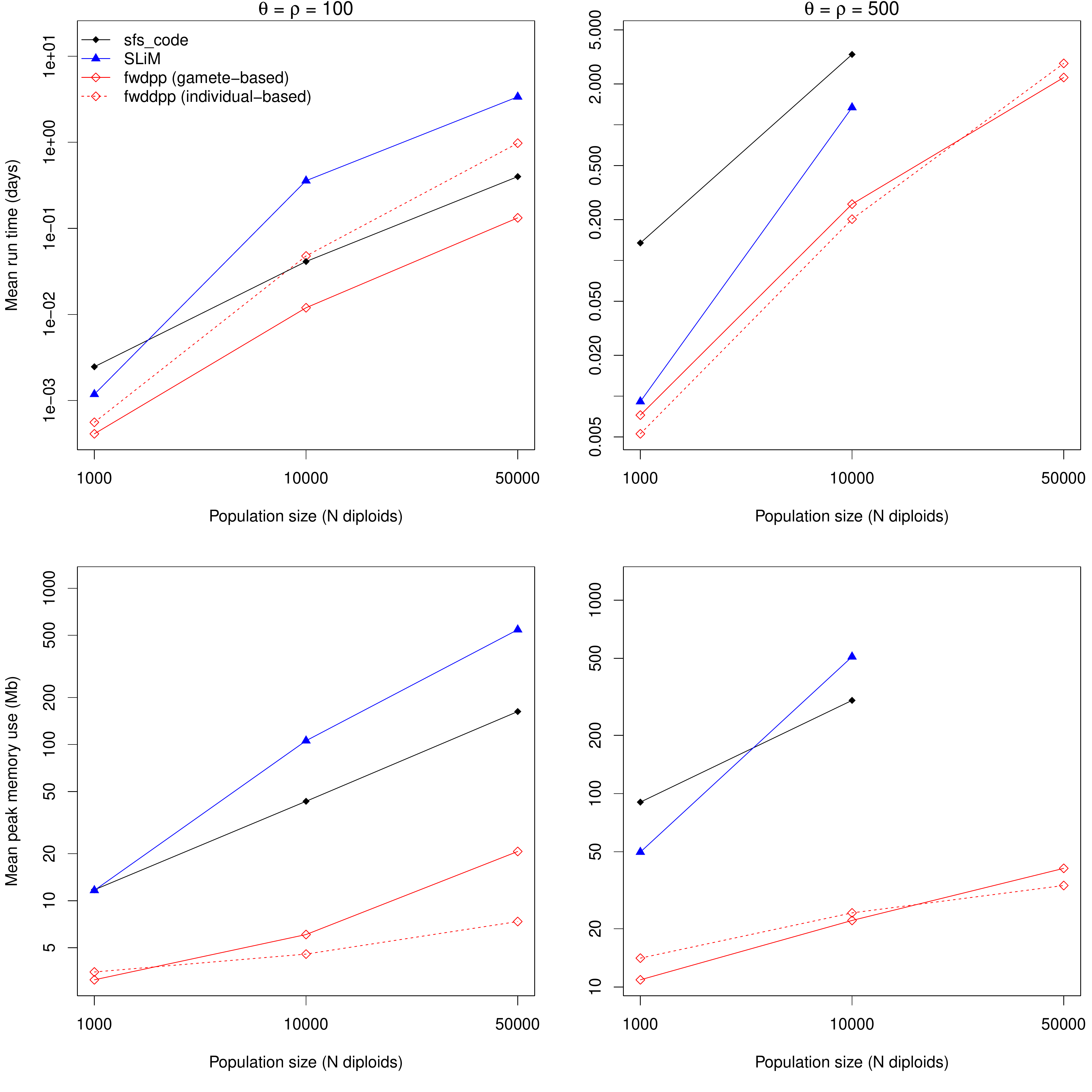}
  \caption{\label{fig:performance2} Performance comparison for the case of large population size ($N$).  Shown are the means of run time and of peak memory use for \texttt{sfs\_code}, \texttt{SLiM}, and a program written using \texttt{fwdpp}.  Note that the y-axis is on a log scale. The left column is for the case of $\theta = \rho = 100$ and the right column shows  $\theta = \rho = 500$ ($\theta$ and $\rho$ refer to the scaled mutation and recombination rates, respectively, for the entire region).   The results are based on 100 replicates of each simulation engine for each value of $N$ and each replicate was evolved for $10N$ generations. Missing data points occurred when a particular simulation did not complete any replicates within seven days, at which point the job was set for automatic termination. For \texttt{SLiM} and \texttt{sfs\_code}, the locus length simulated was $L=10^5$ base pairs and the per-site mutation and recombination rates were chosen to obtain the desired $\theta$ and $\rho$ for the entire region. }
\end{figure}

As $N$ becomes larger, \texttt{fwdpp} becomes much faster than either \texttt{sfs\_code} or \texttt{SLiM} (Figure \ref{fig:performance2}). For populations as large as $N=50,000$ diploids and $\theta = \rho = 100$, \texttt{fwdpp} and \texttt{sfs\_code} were comparable in performance and both were substantially faster than \texttt{SLiM} as $N$ increased.  For $\theta = \rho = 500$, \texttt{fwdpp} was orders of magnitude faster than either \texttt{SLiM} or \texttt{sfs\_code}.

\begin{figure}[!h]
  \centering
\includegraphics[scale=0.6]{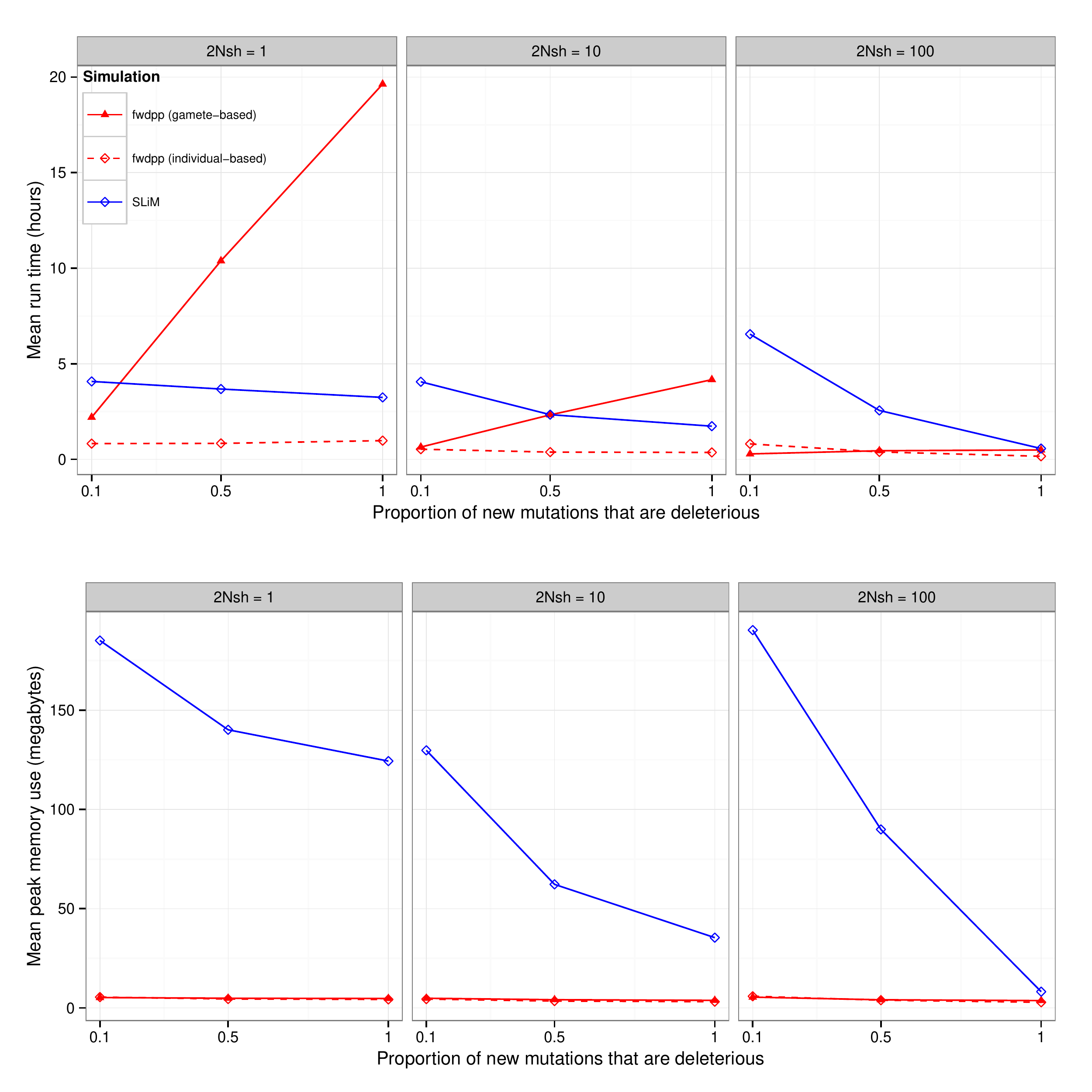}
  \caption{\label{fig:selection}Performance comparison between \texttt{SLiM}, \texttt{fwdpp}'s gamete-based sampling scheme, and \texttt{fwdpp}'s individual-based scheme for models involving both neutral and codominant deleterious alleles.  All results are based on 100 replicates with $N=10^4$ and $10N$ generations of evolution simulated per replicate.  The total mutation rate was chosen such that $2N\mu = 200$ and the proportion of newly-arising deleterious mutations was varied.  The three different panels represent three different strengths of selection against heterozygotes ($2Nsh = 1$, 10, or 100). }
\end{figure}

\begin{figure}[!h]
  \centering
    \begin{subfigure}[b]{0.3\textwidth}
  \centering
  \includegraphics[scale=0.3]{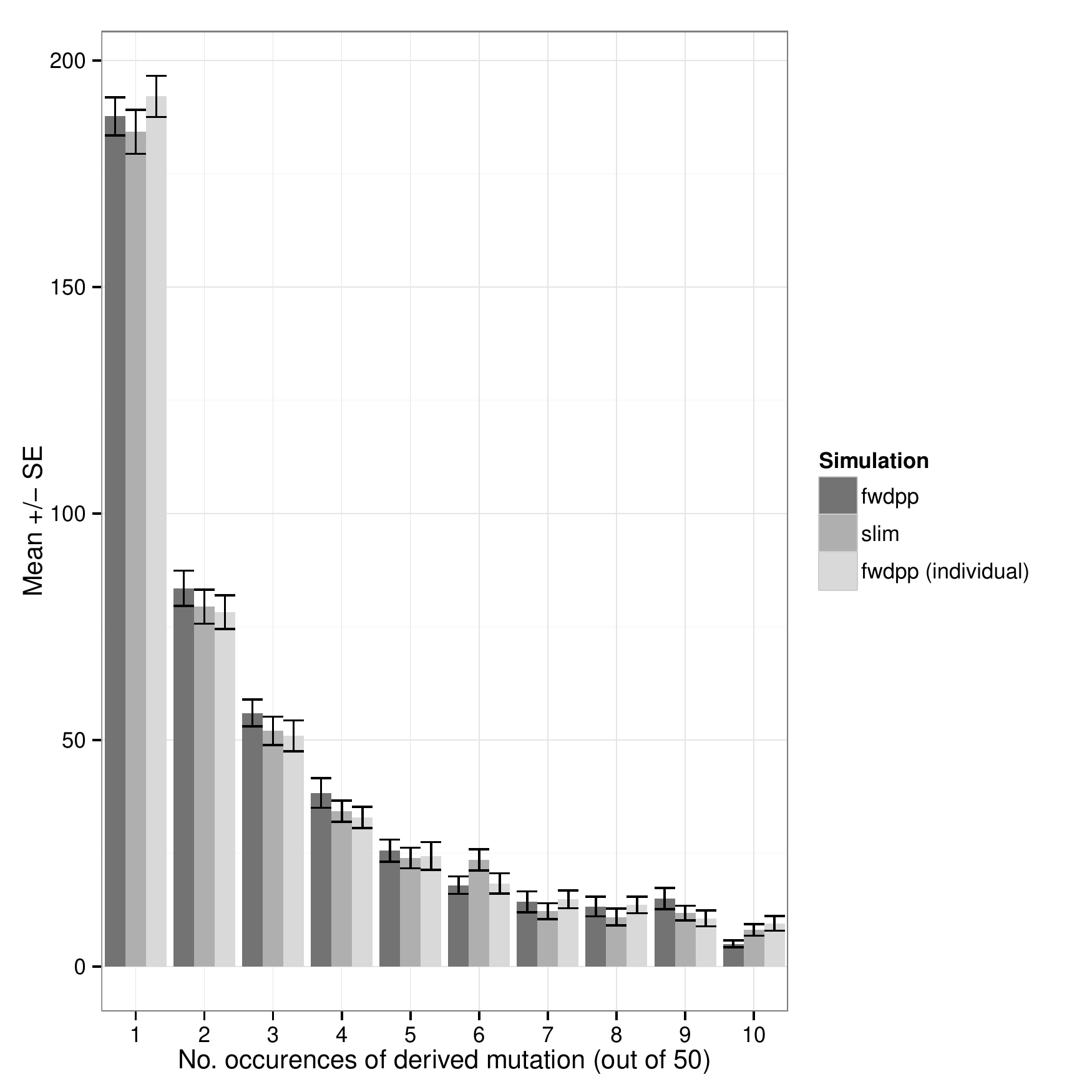}
  \caption{\label{sfig:case3}$2Nsh=1, p = 1$}
  \end{subfigure}
    \begin{subfigure}[b]{0.3\textwidth}
  \centering
  \includegraphics[scale=0.3]{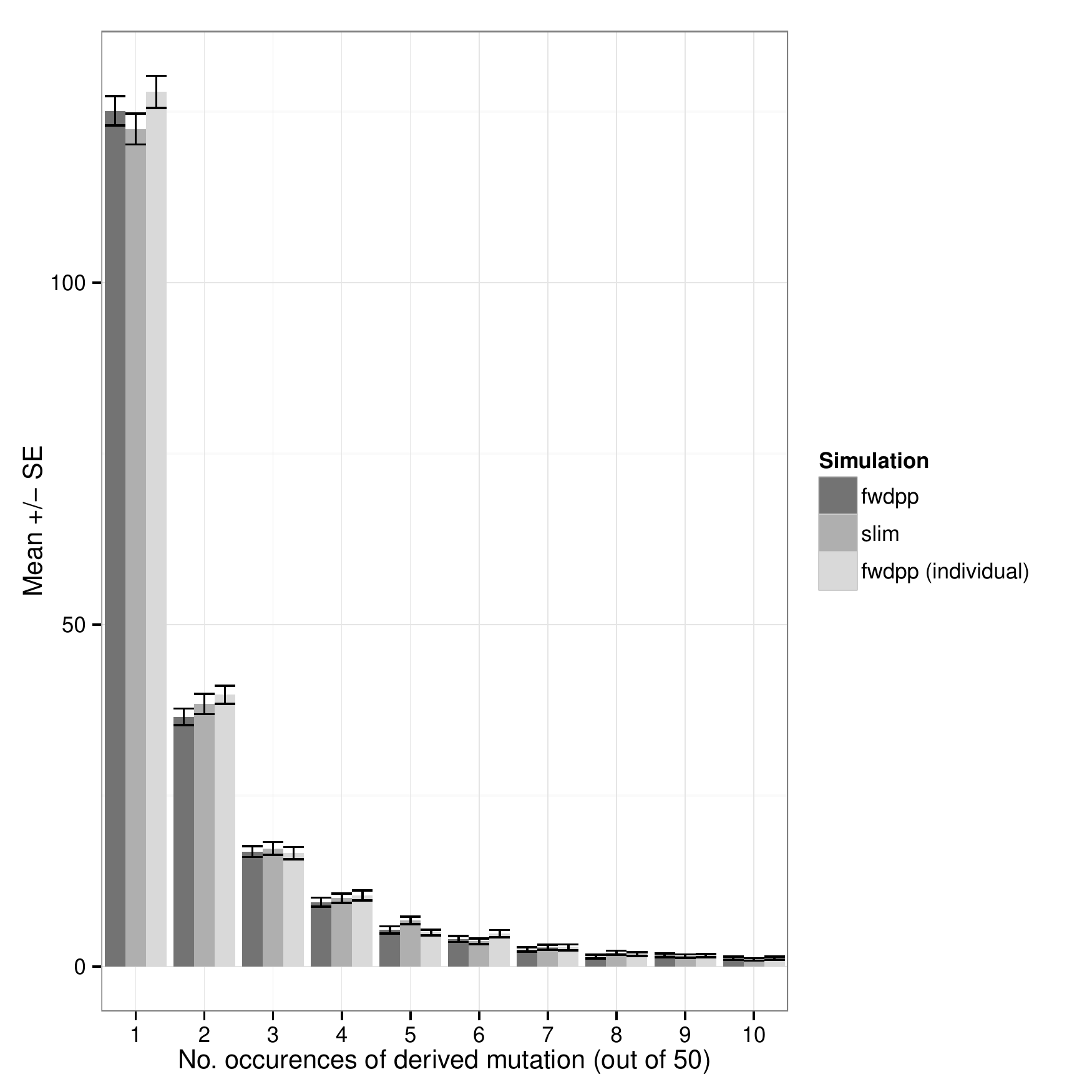}
  \caption{\label{sfig:case6}$2Nsh=10, p = 1$}
  \end{subfigure}
    \begin{subfigure}[b]{0.3\textwidth}
  \centering
  \includegraphics[scale=0.3]{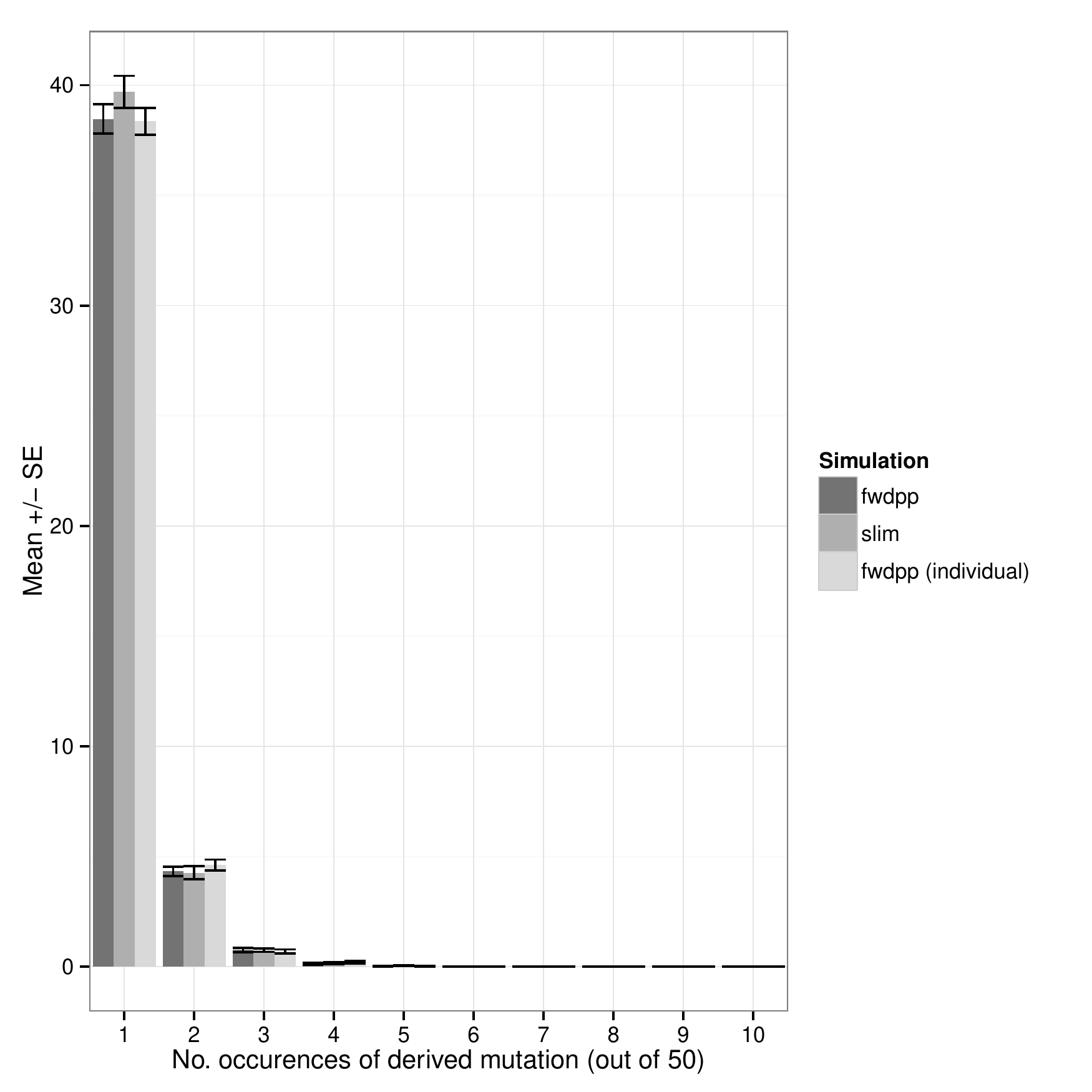}
  \caption{\label{sfig:case9}$2Nsh=100, p = 1$}
  \end{subfigure}
  \caption{\label{fig:selsfs}Site frequency spectra for models with codominant deleterious alleles.  Plots are based on a sample of size $n = 50$ taken from the simulations in Figure \ref{fig:selection} where the proportion of newly-arising deleterious mutations ($p$) was one.}
\end{figure}

The results in Figures \ref{fig:performance1} and \ref{fig:performance2} only consider neutral mutations.  However, coalescent simulations \citep{Hudson:2002vy,Chen:2009vt} should generally be the preferred choice for neutral models because such simulations will typically be much faster than even the fastest forward simulation.  For forward simulations, both the strength of selection and the proportion of selected mutations in the population will affect performance.  Figure \ref{fig:selection} compares the run times and peak memory usage of \texttt{fwdpp} and \texttt{SLiM} for the simple case of selection against codominant mutations with a fixed effect on fitness and multiplicative fitness across sites.  Further, comparison to \texttt{SLiM} seems relevant because it is an efficient and relatively easy to use program that is likely to be widely-used for population-genetic simulations of models with selection. Because \texttt{SLiM} and the example programs written using \texttt{fwdpp} scale fitness differently ($1, 1+sh, 1+s$ and $1, 1+sh, 1+2s$, respectively, for the three genotypes), I chose $s$ and $h$ for each program such that the strength of selection on the three genotypes was the same.  The population size was set to $N = 10^4$ diploids and the total mutation rate was chosen such that $2N\mu = 200$.  The recombination rate was set to 0, and $p$, the proportion of newly-arising mutations that are deleterious, was set to 0.1, 0.5, or 1.  For each value of $p$, 100 replicates were simulated for $10N$ generations.  As $p$ increases and selection gets weaker ($2Nsh$ gets smaller), \texttt{fwdpp}'s gamete-based algorithm gets slower (Figure \ref{fig:selection}).  The case of $2Nsh=1$ and $p= 0.5$ or $1$ is particularly pathological for \texttt{fwdpp}.  However, this parameter combination models a situation where 50\% or 100\% of newly-arising mutations are deleterious with $sh = -\frac{1}{2N}$, and thus selection and drift are comparable in their effects on levels of variation.  In practice, many models of interest will incorporate a distribution of selection coefficients such that this particular case should be viewed as extreme.  For \texttt{SLiM}, the parameters have the opposite effect on performance; \texttt{slim} slows down as selection gets stronger and there are fewer selected mutations in the population.  However, with the exception of the pathological case of a large proportion of weakly-selected mutations, \texttt{SLiM} and \texttt{fwdpp}'s gamete-based sampling scheme showed similar mean run times overall, suggesting that both are capable of efficiently simulating large regions with a substantial fraction of selected mutations and when selection is a stronger force than drift.  For all parameters shown in Figure \ref{fig:selection}, \texttt{fwdpp}'s individual-based sampling method is much more uniform in average run time, typically outperforming both \texttt{SLiM} and \texttt{fwdpp}'s gamete-based method.  As seen in Figures \ref{fig:performance1} and \ref{fig:performance2} above for the case of neutral models, \texttt{fwdpp} uses much less memory than \texttt{SLiM} for models with selection (Figure \ref{fig:selection}).  Finally, Figure \ref{fig:selsfs} shows that \texttt{SLiM} and the two sampling algorithms \texttt{fwdpp} result in nearly-identical deleterious mutation frequencies for the models shown in Figure \ref{fig:selection}, implying that all three methods are of similar accuracy for multi-site models with selection.  The results in Figure \ref{fig:selection} strongly argue that the individual-based sampling routines of \texttt{fwdpp} should be preferred for models involving natural selection.

\section*{Applications}
In this section, I compare the output of programs written using the gamete-based sampler in \texttt{fwdpp} to both theoretical predictions and the output of well-validated coalescent simulations.  Each of the models below is implemented in an example program distributed with the \texttt{fwdpp} code.  For results based on forward simulations, the population size was $N=10^4$ diploids and the sample size taken at the end of the simulation was $n=50$ (from each population in the case of multi-population models).  All summary statistics were calculated using routines from \texttt{libsequence} \citep{Thornton:2003vz}.  For all neutral models, the coalescent simulation program used was \texttt{ms} \citep{Hudson:2002vy}.   The neutral mutation rate and the recombination rate are per region and the region is assumed to be autosomal.  These assumptions result in the scaled mutation rate $\theta=4N\mu$, where $\mu$ is the mutation rate to neutral mutations per gamete per generation, and the scaled recombination rate $\rho = 4Nr$, where $r$ is the probability of crossing over per diploid per generation within the region.  All simulation results are based on 1,000 replicates each of forward and coalescent simulation. 

\subsection*{The equilibrium Wright-Fisher model}
\begin{figure}[!h]
  \centering
  \includegraphics[scale=0.6]{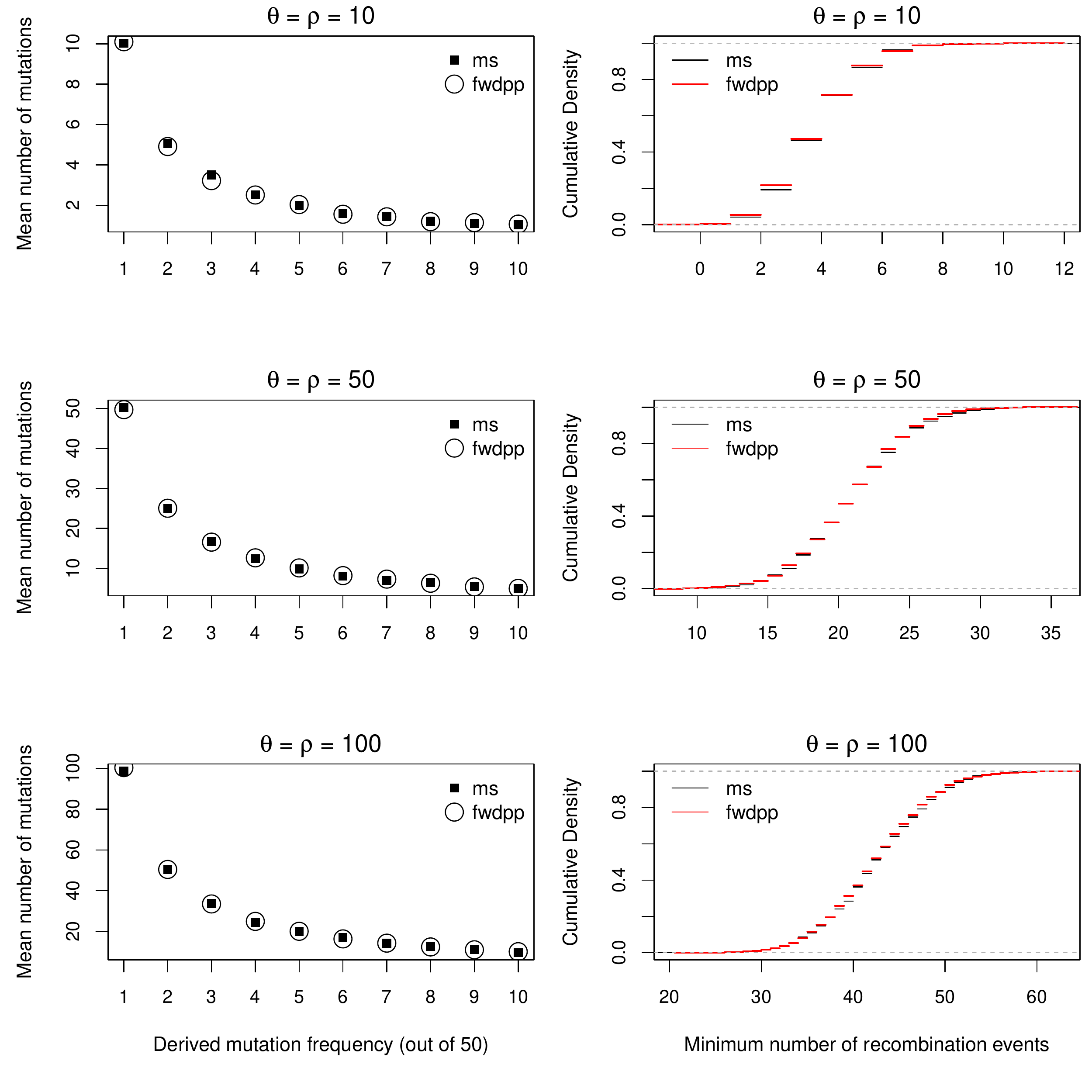}
  \caption{\label{fig:compare2ms}The average site frequency spectrum (left column) and the distribution of the minimum number of recombination events \citep[right column]{Hudson:1985wh} are compared between \texttt{fwdpp} and the coalescent simulation program \texttt{ms} \citep{Hudson:2002vy}.  All results are based on 1,000 simulated replicates.  The forward simulation involved a diploid population of size $N=10^4$ evolving with mutation and recombination occurring at rates $\theta$ and $\rho$, respectively, for $10N$ generations. All summary statistics are based on a sample of size $n = 50$ and were calculated using \texttt{libsequence} \citep{Thornton:2003vz}.}
\end{figure}

We first consider the standard Wright-Fisher model of a constant population and no selection. I performed simulations for each of three parameter values ($\theta=\rho = 10, 50$, and $100$).  Figure \ref{fig:compare2ms} shows the first 10 bins of the site frequency spectrum and the distribution of minimum number of recombination events \citep{Hudson:1985wh} obtained using both simulation methods.  The forward simulation and the coalescent simulation gave identical results (to within Monte Carlo error) in all cases, and there were no significant differences in the distributions of these statistics (Kolmogorov-Smirnov tests, all $P > 0.05$).  All of the results below are based on the gamete-based portion of \texttt{fwdpp} as it is more efficient for models without selection.

\subsection*{Population split followed by equilibrium migration}
\begin{figure}[!h]
  \centering
  \begin{subfigure}[b]{\textwidth}
  \centering
  \includegraphics[scale=2]{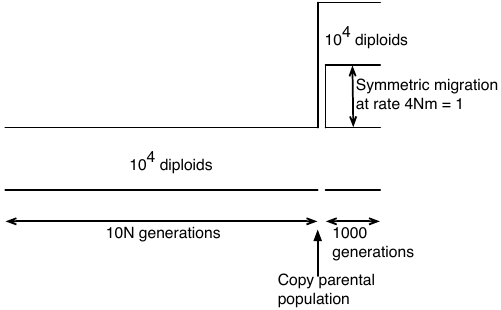}
  \caption{\label{sfig:split}Demographic model}
  \end{subfigure}
  \begin{subfigure}[b]{0.3\textwidth}
    \centering
    \includegraphics[scale=0.3]{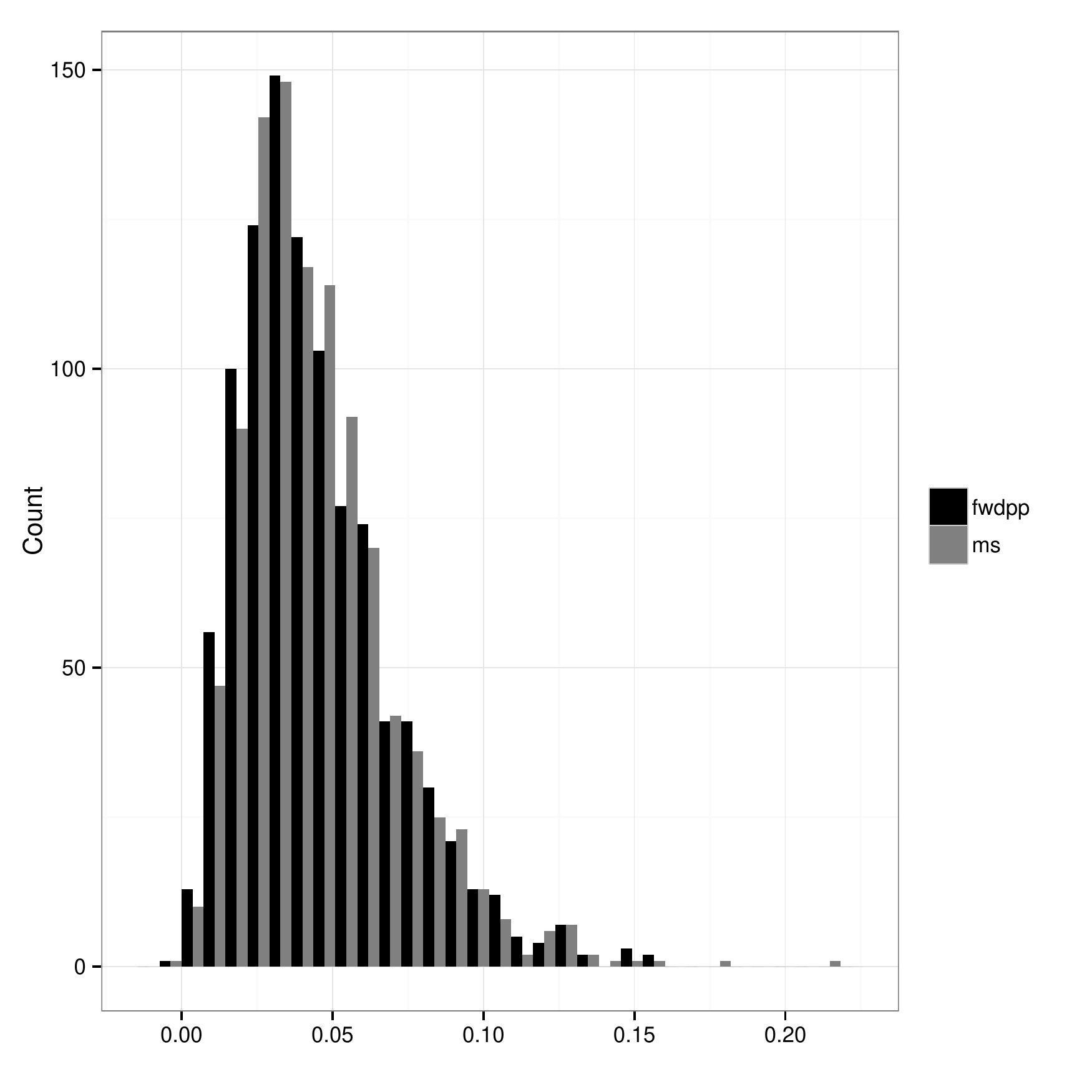}
    \caption{\label{sfig:fst}$F_{ST}$}
  \end{subfigure}
  \begin{subfigure}[b]{0.3\textwidth}
  \centering
    \includegraphics[scale=0.3]{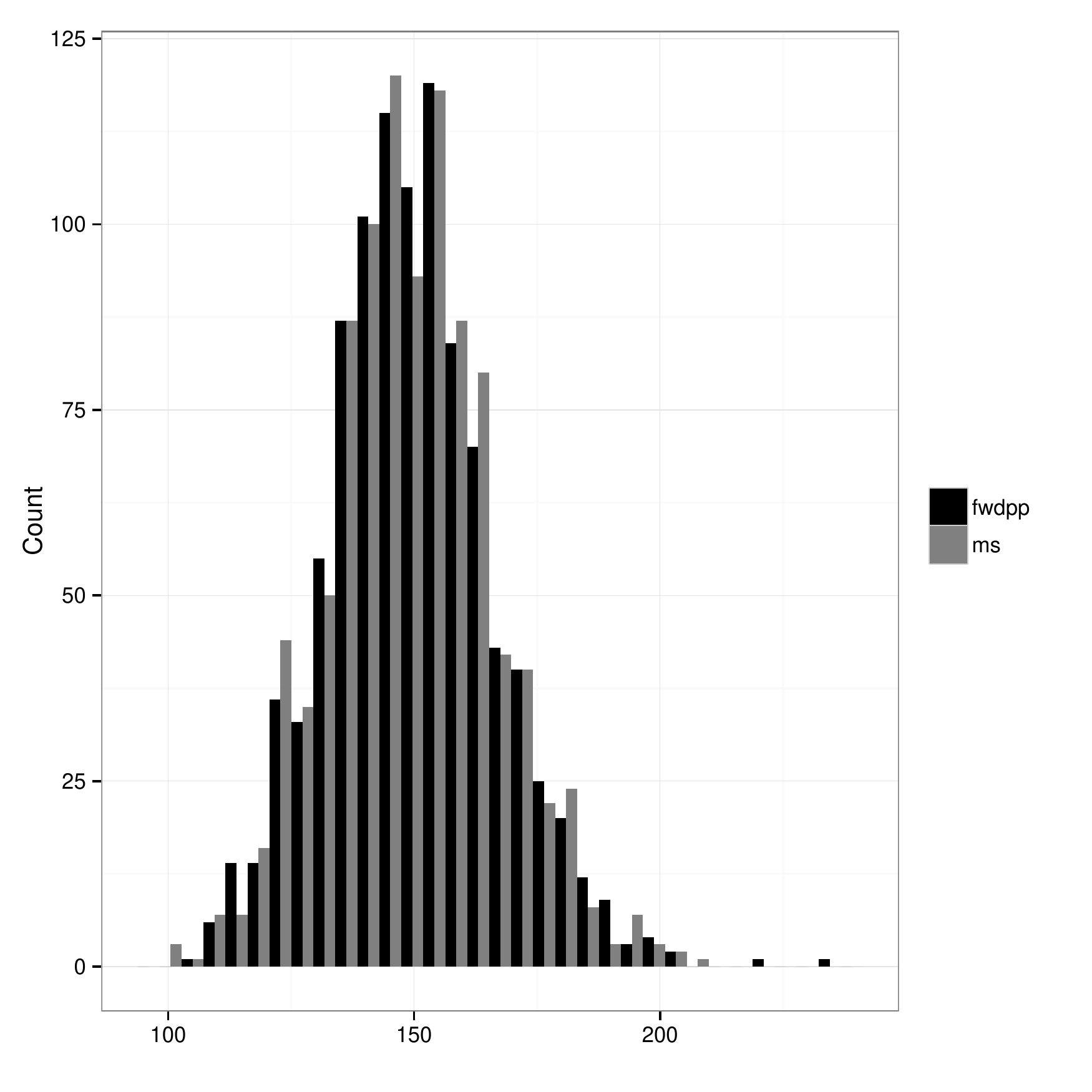}
    \caption{\label{sfig:private}Private polymorphisms}
  \end{subfigure}
  \begin{subfigure}[b]{0.3\textwidth}
  \centering
    \includegraphics[scale=0.3]{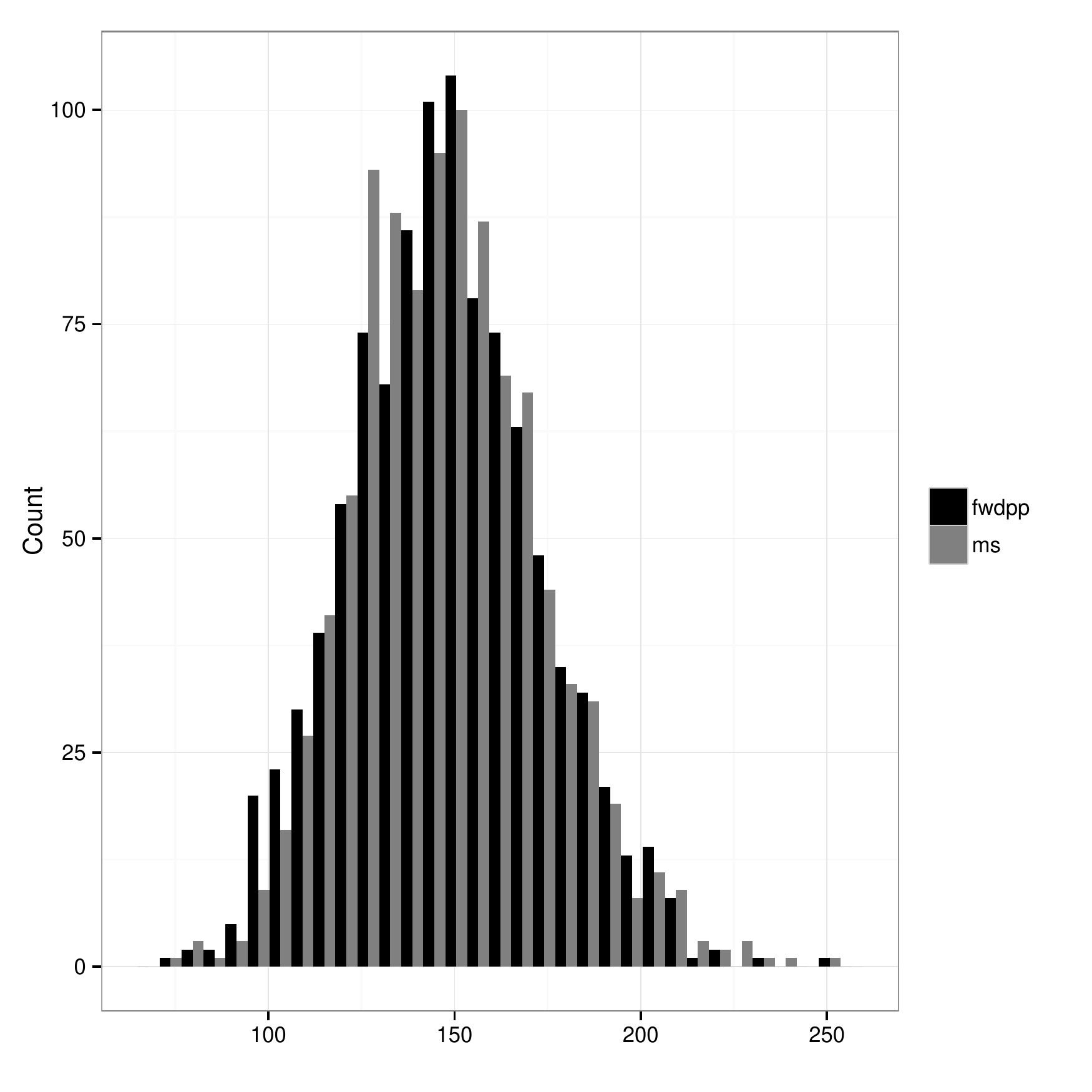}
    \caption{\label{sfig:shared}Shared polymorphisms}
  \end{subfigure}
  \caption{\label{fig:migration}Distributions of genetic variation between populations simulated under a model of recent divergence with migration.  [\ref{sfig:split}] A population split followed by symmetric migration.  An ancestral population of size $N=10^4$ diploids was evolved for $10N$ generations with mutation rate $\theta=50$ and recombination rate $\rho = 50$.  The ancestral population was then split into two equal-sized daughter populations of size $10^4$ (thus resulting in a population split with no bottleneck).  The two populations were evolved for another 1,000 generations with symmetric migration at rate $4Nm = 1$. Figures \ref{sfig:fst}-\ref{sfig:shared} compare results based on 1,000 replicates of forward simulation using \texttt{fwdpp} and 1,000 replicates of the coalescent simulation \texttt{ms} \citep{Hudson:2002vy}. [\ref{sfig:fst}] The distribution of $F_{ST}$ \citep{Hudson:1992wq}. [\ref{sfig:private}] The distribution of the total number of private polymorphisms.  [\ref{sfig:shared}] The distribution of the number of polymorphisms shared between the two populations.}
\end{figure}

I simulated the demographic model shown in Figure \ref{sfig:split} using a forward simulation implemented with \texttt{fwdpp}.  The model in Figure \ref{sfig:split} is equivalent to the following command using the coalescent simulation program \texttt{ms} \citep{Hudson:2002vy}:

\begin{verbatim}
ms 100 1000 -t 50 -r 50 1000 -I 2 50 50 1 -ej 0.025 2 1 -em 0.025 1 2 0.
\end{verbatim}

Figures \ref{sfig:fst}-\ref{sfig:shared} compare the distributions of several summaries of within- and between- population variation.  The forward and coalescent simulations are in excellent agreement, and no significant differences in the distribution of these summary statistic exists (Kolmogorov-Smirnov test, all $P > 0.05$).

\section*{Discussion}
I have described \texttt{fwdpp}, which is a C++ template library designed to help implement efficient forward-time simulations of large populations.  The library's performance compares favorably to other existing simulation engines and has the additional advantage of allowing novel models to be rapidly implemented.  I expect \texttt{fwdpp} to be of particular use when very large samples with selected mutations must be simulated, such as case/control samples or large population-genomic data sets.  The library is under active development and future releases will likely both improve performance as well as add new features.

Importantly, users of forward simulations should appreciate that there may be no single software solution that is ideal for all purposes.  For example, users wishing to evaluate the population-genetic properties of relatively small samples (say $n \leq 100$) under standard population genetic fitness models would perhaps be better-served by \texttt{SLiM} or \texttt{sfs\_code}, as such scenarios can be simulated effectively with either program in reasonable time (Figure \ref{fig:performance1} and  \cite{Messer:2013ct}) by keeping the population size ($N$) small.  Further, \texttt{SLiM} or \texttt{sfs\_code} already implement a variety of relevant demographic models such as migration and changing population size. The intent of \texttt{fwdpp} is to offer a combination of modeling flexibility and speed not currently found in existing forward simulation programs and to provide a library interface to that flexibility.  There are several scenarios where \texttt{fwdpp} may be the preferred tool.  First, for models requiring large $N$ and selection, \texttt{fwdpp} may be the fastest algorithm (Figures \ref{fig:performance2} and \ref{fig:selection}).  Second, when non-standard fitness models and/or phenotype-to-fitness relationships are required (such as in \cite{thornton:2013dl}), \texttt{fwdpp} provides a flexible system for implementing such models while also allowing for complex demographics, complementing existing efforts in this area \citep{Kessner:ip,Neuenschwander:2008bf,Peng:2008em,Pinelli:2012fc}.  Finally, \texttt{fwdpp} is likely to be useful when the user needs to maximize run-time efficiency for a particular demographic scenario and does not require the flexibility of a more general program.  

\section*{Acknowledgements}
This work was funded by NIH grant GM085183 to KRT.  I thanks Jeffrey Ross-Ibarra for helpful comments on the manuscript and Ryan Hernandez for discussion about, and valuable assistance with, \texttt{sfs\_code}.  I also thank two anonymous reviewers whose comments greatly improved the manuscript.

\bibliography{references}

\begin{thebibliography}{29}
\expandafter\ifx\csname natexlab\endcsname\relax\def\natexlab#1{#1}\fi

\bibitem[{{\sc {1000 Genomes Project Consortium}} {\em et~al.\/}(2010){\sc
  {1000 Genomes Project Consortium}}, {\sc Abecasis}, {\sc Altshuler}, {\sc
  Auton}, {\sc Brooks} {\em et~al.\/}}]{GenomesProjectConsortium:2010gj}
{\sc {1000 Genomes Project Consortium}}, {\sc G.~R. Abecasis}, {\sc
  D.~Altshuler}, {\sc A.~Auton}, {\sc L.~D. Brooks} {\em et~al.\/}, 2010 {A map
  of human genome variation from population-scale sequencing.}
\newblock Nature {\bf 467}: 1061--1073.

\bibitem[{{\sc Baldwin-Brown} {\em et~al.\/}(2014){\sc Baldwin-Brown}, {\sc
  Long} and {\sc Thornton}}]{BaldwinBrown:2014eaa}
{\sc Baldwin-Brown, J.~G.}, {\sc A.~D. Long} and {\sc K.~R. Thornton}, 2014
  {The Power to Detect Quantitative Trait Loci Using Resequenced,
  Experimentally Evolved Populations of Diploid, Sexual Organisms}.
\newblock Molecular Biology and Evolution {\bf 31}: 1040--1055.

\bibitem[{{\sc Burke} {\em et~al.\/}(2010){\sc Burke}, {\sc Dunham}, {\sc
  Shahrestani}, {\sc Thornton}, {\sc Rose} {\em et~al.\/}}]{Burke:2010eq}
{\sc Burke, M.~K.}, {\sc J.~P. Dunham}, {\sc P.~Shahrestani}, {\sc K.~R.
  Thornton}, {\sc M.~R. Rose} {\em et~al.\/}, 2010 {Genome-wide analysis of a
  long-term evolution experiment with Drosophila.}
\newblock Nature {\bf 467}: 587--590.

\bibitem[{{\sc Burton} {\em et~al.\/}(2007){\sc Burton}, {\sc Clayton}, {\sc
  Cardon}, {\sc Craddock}, {\sc Deloukas} {\em et~al.\/}}]{Burton:2007ht}
{\sc Burton, P.~R.}, {\sc D.~G. Clayton}, {\sc L.~R. Cardon}, {\sc
  N.~Craddock}, {\sc P.~Deloukas} {\em et~al.\/}, 2007 {Genome-wide association
  study of 14,000 cases of seven common diseases and 3,000 shared controls}.
\newblock Nature {\bf 447}: 661--678.

\bibitem[{{\sc Cao} {\em et~al.\/}(2011){\sc Cao}, {\sc Schneeberger}, {\sc
  Ossowski}, {\sc G{\"u}nther}, {\sc Bender} {\em et~al.\/}}]{Cao:2011cfa}
{\sc Cao, J.}, {\sc K.~Schneeberger}, {\sc S.~Ossowski}, {\sc T.~G{\"u}nther},
  {\sc S.~Bender} {\em et~al.\/}, 2011 {Whole-genome sequencing of multiple
  Arabidopsis thaliana populations}.
\newblock Nature Genetics {\bf 43}: 956--963.

\bibitem[{{\sc Carvajal-Rodr{\'\i}guez}(2008)}]{CarvajalRodriguez:2008gz}
{\sc Carvajal-Rodr{\'\i}guez, A.}, 2008 {GENOMEPOP: a program to simulate
  genomes in populations.}
\newblock BMC Bioinformatics {\bf 9}: 223.

\bibitem[{{\sc Chadeau-Hyam} {\em et~al.\/}(2008){\sc Chadeau-Hyam}, {\sc
  Hoggart}, {\sc O'reilly}, {\sc Whittaker}, {\sc De~Iorio} {\em
  et~al.\/}}]{ChadeauHyam:2008gj}
{\sc Chadeau-Hyam, M.}, {\sc C.~J. Hoggart}, {\sc P.~F. O'reilly}, {\sc J.~C.
  Whittaker}, {\sc M.~De~Iorio} {\em et~al.\/}, 2008 {Fregene: simulation of
  realistic sequence-level data in populations and ascertained samples}.
\newblock BMC Bioinformatics {\bf 9}: 364.

\bibitem[{{\sc Chen} {\em et~al.\/}(2009){\sc Chen}, {\sc Marjoram} and {\sc
  Wall}}]{Chen:2009vt}
{\sc Chen, G.~K.}, {\sc P.~Marjoram} and {\sc J.~D. Wall}, 2009 {Fast and
  flexible simulation of DNA sequence data}.
\newblock Genome Research {\bf 19}: 136--142.

\bibitem[{{\sc Crow} and {\sc Kimura}(1971)}]{Crow:1971wa}
{\sc Crow, J.~F.} and {\sc M.~Kimura}, 1971 {\em {An Introduction to Population
  Genetics Theory}\/}.
\newblock Alpha Editions.

\bibitem[{{\sc Hernandez}(2008)}]{Hernandez:2008kn}
{\sc Hernandez, R.~D.}, 2008 {A flexible forward simulator for populations
  subject to selection and demography}.
\newblock Bioinformatics (Oxford, England) {\bf 24}: 2786--2787.

\bibitem[{{\sc Hoggart} {\em et~al.\/}(2007){\sc Hoggart}, {\sc Chadeau-Hyam},
  {\sc Clark}, {\sc Lampariello}, {\sc Whittaker} {\em
  et~al.\/}}]{Hoggart:2007kp}
{\sc Hoggart, C.~J.}, {\sc M.~Chadeau-Hyam}, {\sc T.~G. Clark}, {\sc
  R.~Lampariello}, {\sc J.~C. Whittaker} {\em et~al.\/}, 2007 {Sequence-Level
  Population Simulations Over Large Genomic Regions}.
\newblock Genetics {\bf 177}: 1725--1731.

\bibitem[{{\sc Hudson}(2002)}]{Hudson:2002vy}
{\sc Hudson, R.~R.}, 2002 {Generating samples under a Wright-Fisher neutral
  model of genetic variation.}
\newblock Bioinformatics (Oxford, England) {\bf 18}: 337--338.

\bibitem[{{\sc Hudson} and {\sc Kaplan}(1985)}]{Hudson:1985wh}
{\sc Hudson, R.~R.} and {\sc N.~L. Kaplan}, 1985 {Statistical properties of the
  number of recombination events in the history of a sample of DNA sequences.}
\newblock Genetics {\bf 111}: 147--164.

\bibitem[{{\sc Hudson} {\em et~al.\/}(1992){\sc Hudson}, {\sc Slatkin} and {\sc
  Maddison}}]{Hudson:1992wq}
{\sc Hudson, R.~R.}, {\sc M.~Slatkin} and {\sc W.~P. Maddison}, 1992
  {Estimation of levels of gene flow from DNA sequence data.}
\newblock Genetics {\bf 132}: 583--589.

\bibitem[{{\sc Josuttis}(1999)}]{Josuttis:1999tx}
{\sc Josuttis, N.}, 1999 {\em {The C++ Standard Library: A Tutorial and
  Reference}\/}.
\newblock Addison-Wesley, first edition.

\bibitem[{{\sc Kessner} and {\sc Novembre}(2014)}]{Kessner:ip}
{\sc Kessner, D.} and {\sc J.~Novembre}, 2014 {forqs: forward-in-time
  simulation of recombination, quantitative traits and selection.}
\newblock Bioinformatics (Oxford, England) {\bf 30}: 576--577.

\bibitem[{{\sc Kimura}(1969)}]{Kimura:1969tn}
{\sc Kimura, M.}, 1969 {The number of heterozygous nucleotide sites maintained
  in a finite population due to steady flux of mutations.}
\newblock Genetics {\bf 61}: 893--903.

\bibitem[{{\sc Mackay} {\em et~al.\/}(2012){\sc Mackay}, {\sc Richards}, {\sc
  Stone}, {\sc Barbadilla}, {\sc Ayroles} {\em et~al.\/}}]{Mackay:2012fda}
{\sc Mackay, T. F.~C.}, {\sc S.~Richards}, {\sc E.~A. Stone}, {\sc
  A.~Barbadilla}, {\sc J.~F. Ayroles} {\em et~al.\/}, 2012 {The Drosophila
  melanogaster Genetic Reference Panel.}
\newblock Nature {\bf 482}: 173--178.

\bibitem[{{\sc McVean} {\em et~al.\/}(2012){\sc McVean}, {\sc Altshuler
  Co-Chair}, {\sc Durbin Co-Chair}, {\sc Abecasis}, {\sc Bentley} {\em
  et~al.\/}}]{McVean:2012co}
{\sc McVean, G.~A.}, {\sc D.~M. Altshuler Co-Chair}, {\sc R.~M. Durbin
  Co-Chair}, {\sc G.~R. Abecasis}, {\sc D.~R. Bentley} {\em et~al.\/}, 2012 {An
  integrated map of genetic variation from 1,092 human genomes}.
\newblock Nature {\bf 491}: 56--65.

\bibitem[{{\sc Messer}(2013)}]{Messer:2013ct}
{\sc Messer, P.~W.}, 2013 {SLiM: simulating evolution with selection and
  linkage.}
\newblock Genetics {\bf 194}: 1037--1039.

\bibitem[{{\sc Neuenschwander} {\em et~al.\/}(2008){\sc Neuenschwander}, {\sc
  Hospital}, {\sc Guillaume} and {\sc Goudet}}]{Neuenschwander:2008bf}
{\sc Neuenschwander, S.}, {\sc F.~Hospital}, {\sc F.~Guillaume} and {\sc
  J.~Goudet}, 2008 {quantiNemo: an individual-based program to simulate
  quantitative traits with explicit genetic architecture in a dynamic
  metapopulation.}
\newblock Bioinformatics (Oxford, England) {\bf 24}: 1552--1553.

\bibitem[{{\sc Padhukasahasram} {\em et~al.\/}(2008){\sc Padhukasahasram}, {\sc
  Marjoram}, {\sc Wall}, {\sc Bustamante} and {\sc
  Nordborg}}]{Padhukasahasram:2008ee}
{\sc Padhukasahasram, B.}, {\sc P.~Marjoram}, {\sc J.~D. Wall}, {\sc C.~D.
  Bustamante} and {\sc M.~Nordborg}, 2008 {Exploring population genetic models
  with recombination using efficient forward-time simulations}.
\newblock Genetics {\bf 178}: 2417--2427.

\bibitem[{{\sc Peng} and {\sc Amos}(2008)}]{Peng:2008em}
{\sc Peng, B.} and {\sc C.~I. Amos}, 2008 {Forward-time simulations of
  non-random mating populations using simuPOP.}
\newblock Bioinformatics (Oxford, England) {\bf 24}: 1408--1409.

\bibitem[{{\sc Peng} {\em et~al.\/}(2007){\sc Peng}, {\sc Amos} and {\sc
  Kimmel}}]{Peng:2007fx}
{\sc Peng, B.}, {\sc C.~I. Amos} and {\sc M.~Kimmel}, 2007 {Forward-time
  simulations of human populations with complex diseases.}
\newblock PLoS Genetics {\bf 3}: e47.

\bibitem[{{\sc Peng} and {\sc Liu}(2010)}]{Peng:2010bi}
{\sc Peng, B.} and {\sc X.~Liu}, 2010 {Simulating sequences of the human genome
  with rare variants}.
\newblock Human Heredity {\bf 70}: 287--291.

\bibitem[{{\sc Pinelli} {\em et~al.\/}(2012){\sc Pinelli}, {\sc Scala}, {\sc
  Amato}, {\sc Cocozza} and {\sc Miele}}]{Pinelli:2012fc}
{\sc Pinelli, M.}, {\sc G.~Scala}, {\sc R.~Amato}, {\sc S.~Cocozza} and {\sc
  G.~Miele}, 2012 {Simulating gene-gene and gene-environment interactions in
  complex diseases: Gene-Environment iNteraction Simulator 2.}
\newblock BMC Bioinformatics {\bf 13}: 132.

\bibitem[{{\sc Pool} {\em et~al.\/}(2012){\sc Pool}, {\sc Corbett-Detig}, {\sc
  Sugino}, {\sc Stevens}, {\sc Cardeno} {\em et~al.\/}}]{Pool:2012cx}
{\sc Pool, J.~E.}, {\sc R.~B. Corbett-Detig}, {\sc R.~P. Sugino}, {\sc K.~A.
  Stevens}, {\sc C.~M. Cardeno} {\em et~al.\/}, 2012 {Population Genomics of
  Sub-Saharan Drosophila melanogaster: African Diversity and Non-African
  Admixture}.
\newblock PLoS Genetics {\bf 8}: e1003080.

\bibitem[{{\sc Thornton}(2003)}]{Thornton:2003vz}
{\sc Thornton, K.}, 2003 {Libsequence: a C++ class library for evolutionary
  genetic analysis.}
\newblock Bioinformatics (Oxford, England) {\bf 19}: 2325--2327.

\bibitem[{{\sc Thornton} {\em et~al.\/}(2013){\sc Thornton}, {\sc Foran} and
  {\sc Long}}]{thornton:2013dl}
{\sc Thornton, K.~R.}, {\sc A.~J. Foran} and {\sc A.~D. Long}, 2013 {Properties
  and Modeling of GWAS when Complex Disease Risk Is Due to Non-Complementing,
  Deleterious Mutations in Genes of Large Effect.}
\newblock PLoS Genetics {\bf 9}: e1003258.

\end{thebibliography}
\bibliographystyle{genetics}

\newpage

\newpage

\newpage

\newpage

\newpage

\newpage

\newpage

%supplementary material
\clearpage

\section*{Supplementary Material}
\renewcommand{\thefigure}{S\arabic{figure}}
\setcounter{figure}{0}

\begin{figure}[!h]
  \centering
  \includegraphics[scale=0.5]{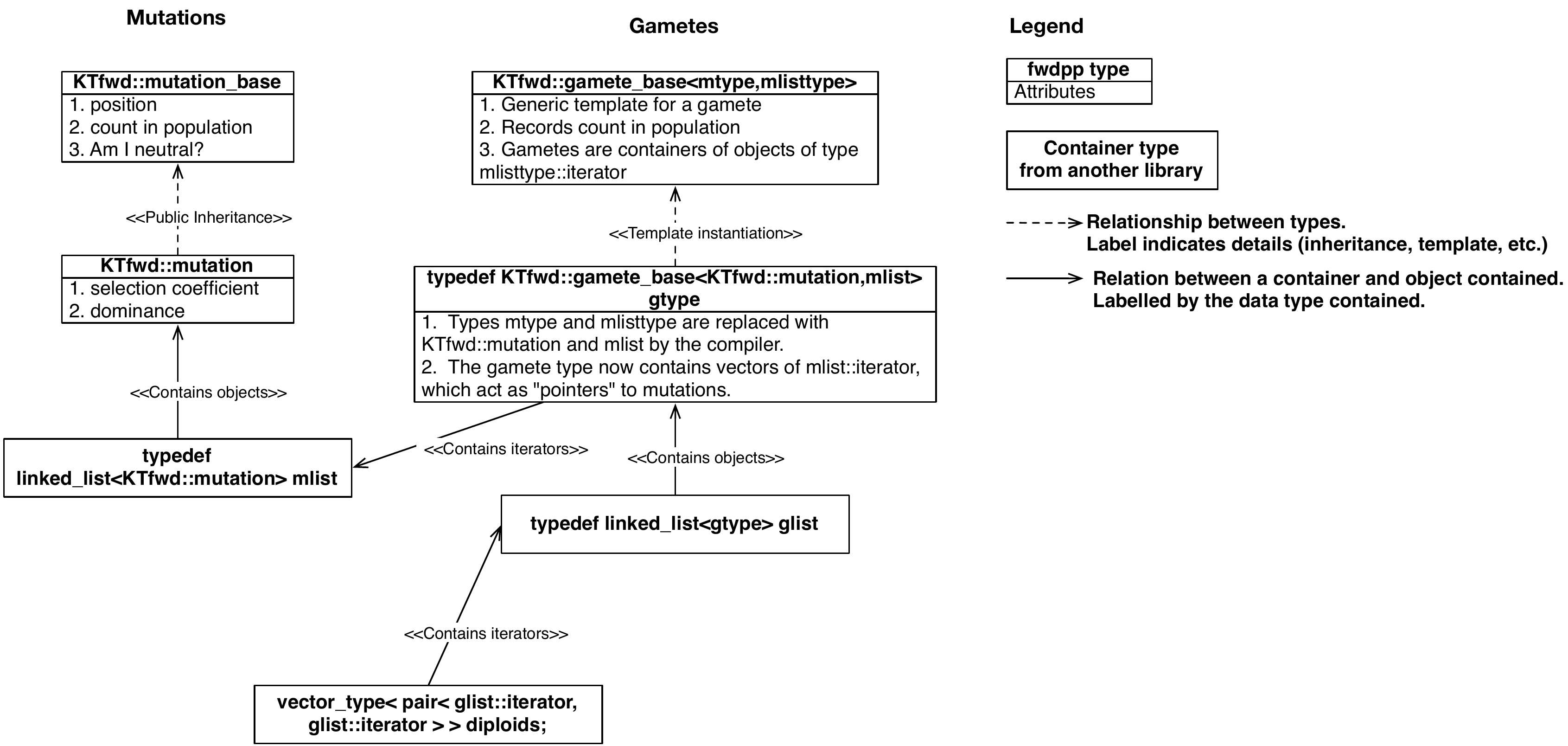}
\caption{\label{fig:uml}Detailed relationships between data types used for individual-based forward simulation.  The first data type defined by the library is \texttt{mutation\_base}, which contains three data attributes (position, number of occurrences, and whether or not the mutation is neutral). The library also provides the data type \texttt{mutation}, which extends \texttt{mutation\_base} by adding selection and dominance values.  Users may provide their own extensions of \texttt{mutation\_base}.  Gametes are implemented using \texttt{gamete\_base}, which is a template data type whose template argument \texttt{mtype} must be \texttt{mutation\_base} or a type derived from \texttt{mutation\_base} (this requirement is enforced at compile-time).  The template argument \texttt{mlisttype} is the type of container in which the simulation will store mutations (this container must be a doubly-linked list).  Gametes are essentially containers of objects called iterators, which point to the mutations carried by the gamete. (In C++, an iterator is a data type abstracting the concept of a pointer \citep[p. 83]{Josuttis:1999tx} and is required because many of the storage containers used in \texttt{fwdpp} are not simple random-access containers.)  In this example, our gamete type is a container of pointers to objects of type \texttt{mutation}.  The collection of mutations and gametes currently in the population are stored in separate doubly-linked lists. The pointers stored by gametes point to elements within the list of mutations.  Finally, diploids are pairs of pointers to gametes and the ``individuals'' in the population are stored in a vector of such pairs.  Because of the template-based implementation of \texttt{fwdpp}, the list and vector types may come either from the C++ standard library \citep[p. 76 and 79]{Josuttis:1999tx} or from any other library providing compatible containers.}
\end{figure}

\end{document}